\documentstyle[12pt]{article}

\begin{document}

\hoffset = -1truecm \voffset = -2truecm \baselineskip = 10 mm

\title{\bf Properties of the gluon recombination functions}

\author{
{\bf Wei Zhu} and {\bf Zhenqi Shen} \\
\normalsize Department of Physics, East China Normal University,
Shanghai 200062, P.R. China \\
}

\date{}

\newpage

\maketitle

\vskip 3truecm

\begin{abstract}

     The gluon recombination functions in the twist-4 QCD evolution
equations are studied at the leading logarithmic approximation
using both the covariant and non-covariant methods.  We justify
that the infrared singularities in the twist-4 coefficient
functions impede us to simply separate the QCD evolution kernels
from the coefficient functions at the equivalent particle
approximation. In particulary, we point out that the gluon
recombination functions in the GLR-MQ evolution equation are
unavailable. The methods avoiding the IR divergences are
discussed, which can be used in the derivations of the evolution
kernels and coefficient functions at the higher twist level.

\end{abstract}

\newpage
\begin{center}
\section{Introduction}
\end{center}

    The QCD evolution equations in the leading order level
predict strong rise of parton densities when the Bjorken variable
$x_B=Q^2/2p.q$ decrease toward small values. This behavior
violates unitarity. Therefore, various models are proposed to
modify the twist-2 evolution kernels (i.e., the parton splitting
functions). Gribov, Levin and Ryskin [1] first suggest a
non-linear evolution equation, in which the evolution kernels (we
call them as the gluon recombination functions) are constructed by
the fan diagrams. Later Mueller and Qiu [2] calculate the (real)
gluon recombination functions at the double leading logarithmic
approximation ($DLLA$) in a covariant perturbation framework. The
GLR-MQ equation is broadly regarded as a key link from
perturbation region to non-perturbation region. This equation was
generalized to include the contributions from more higher order
corrections in the Glauber-Mueller formula [3]. Following them, a
similar evolution equation (we call it as the recombination
equation) is derived in a more broad kinematic region (the leading
logarithmic ($Q^2$) approximation-$LLA(Q^2)$) in Ref.[4].
Different from the GLR-MQ equation, the gluon recombination
functions in this equation are calculated and summed in the
(old-fashioned) time-ordered perturbation theory (TOPT) [5].

    In this paper we present the different results of the gluon
recombination functions in the above mentioned two evolution
equations. The cross section (or structure functions) of deep
inelastic electron-proton scattering in a factorization scheme can
be factorized into the convolution of coefficient functions with
nonperturbative matrix elements, which are defined as the parton
distributions in the twist-2 level. The QCD evolution kernels (for
example, splitting and recombination functions) are separated from
the coefficient functions either in a covariant perturbation
theory or in a parton model using TOPT.  It is well known that
these two approaches are completely equivalent. However, we find
that the infrared (IR) singularities in the twist-4 Feynman
diagrams impede us to safely extract the evolution kernels
$before$ these singularities are cancelled. This result leads to
the fact that the gluon recombination functions in the GLR-MQ
evolution equation are unavailable. For illustrating our idea in
detail, in Sec. 2 we compare two approaches in the derivations of
a splinting function in the twist-2 DGLAP equation [6]. Then we
show two different derivations of the twist-4 recombination
functions in Sec. 3. In Sec. 4  we analyze the IR singularities in
the recombination functions, which arise the questions in the
derivations of the gluon recombination functions in the covariant
theory. The discussions and conclusions are given in Sec. 5.

\newpage
\begin{center}
\section{Twist-2 splitting functions}
\end{center}

   The splitting functions for partons (quarks or gluon) in
the twist-2 evolution equations are schematically defined as Fig.
1a. We emphasize that all initial and final partons in Fig.1
should be on-shell, thus the fraction of the momentum and
helicity, which is carried by a parton, can be fixed [6]. The
physical gauge is convenient in the calculations of the evolution
kernels. In this work we take the physical axial gauge and let the
light-like vector $n$ to fix the gauge as $n\cdot A=0$, $A$ being
the gluon field, where

$$n^{\mu}\equiv \frac{1}{\sqrt{2}}(1,0_{\perp},-1),$$
and

$$\overline{n}^{\mu}\equiv \frac{1}{\sqrt{2}}(1,0_{\perp},1). \eqno(2.1) $$
Therefore, we have

$$n\cdot n=\overline{n}\cdot \overline{n}=0,~~\overline{n}\cdot n=1.\eqno(2.2)$$
Of cause, we can also use other gauges since the result is
gauge-invariant. However, physical picture of a DIS process is
frame- and gauge-dependent. An appropriate choice of the
coordinate frame and gauge is helpful to extract the evolution
kernels using a simple partonic picture.

     A natural derivation of the splitting function is given by
TOPT in a special infinite momentum frame, i.e., the Bjorken
frame, where the virtual photon has almost zero energy $q_0=
(2M\nu+Q^2)/4P$ and longitudinal momentum $q_3=-(2M\nu-Q^2)/4P$. A
Feynman propagator is decomposed into the forward and backward
propagators in TOPT, where the propagating partons stay on-shell,
respectively. One can find that the contributions of the backward
components of the propagators in the cut diagram, for example in
Fig. 2a are suppressed [4]. Thus, the splitting function can be
isolated in the equivalent particle approximation [7] as shown in
Fig. 2b. To illustrate it, we parameterize the momenta in Fig. 2a
as

$$p=(p_0, p_T, p_3)=(p_z, 0, p_z),$$

$$\hat{k}\simeq (zp_z+\frac{k_{\perp}^2}{2zp_z}, k_{\perp}, zp_z),$$

$$k_2\simeq ((1-z)p_z+\frac{k_{\perp}^2}{2(1-z)p_z}, -k_{\perp}, (1-z)p_z). \eqno(2.3)$$

     The contributions of Fig. 2a to the twist-2 coefficient functions are

\newcommand{\slashl}[1]{#1\hspace{-0.59em}/}

$$C_{TOPT}^{twist-2}=\int \frac{d^3\vec{k}_2}{(2\pi)^3}\frac{1}{8E_{\hat{k}}E_{k_2}}\left [\frac{1}{E_{\hat{k}}+E_{k_2}-E_p}
\right ]^2\vert \overline{M}(q\rightarrow qg)\vert^2$$
$$\times \frac{E_p}{E_{\hat{k}}}\overline{M}(\gamma^*q\rightarrow\gamma^*q)\frac{x_B}{Q^2}\delta(z-x_B), \eqno(2.4)$$
where we used

$$Tr[\gamma_{\mu}\slashl{k}_1\gamma_{\nu}
\slashl{\hat{k}}\gamma_{\beta}\slashl{k}_2\gamma_{\alpha}\slashl{\hat{k}}]$$
$$=Tr[\gamma_{\mu}\slashl{k}_1\gamma_{\nu} \slashl{\hat{k}}]Tr[\gamma_{\beta}\slashl{k}_2
\gamma_{\alpha}\slashl{\hat{k}}], \eqno(2.5)$$ since the momentum
$\hat{k}$ is on-shell. Now we can separately define a splitting
function $P_{TOPT}^{q\rightarrow q}$ (Fig. 2b) as

$$\frac{\alpha_s}{2\pi}\frac{dk^2_T}{k^2_T}P_{TOPT}^{q\rightarrow q}$$
$$=\frac{1}{8E_{\hat{k}}E_{k_2}}\left [\frac{1}{E_{\hat{k}}+E_{k_2}-E_p}\right ]^2
\vert \overline{M}(q\rightarrow
q)\vert^2\frac{d^3\vec{k}_2}{(2\pi)^3}, \eqno(2.6)$$ i.e.,

$$P_{TOPT}^{q\rightarrow q}=\frac{4}{3}\left(\frac{1+z^2}{1-z}\right). \eqno(2.7)$$

     The same splitting function can be obtained in a covariant
framework [8]. The related dominant Fynmann diagram for the
coefficient function at the $LLA(Q^2)$ is shown in Fig. 3. For
convenience, we take a collinear (infinite momentum) frame, in
which the momenta of the initial parton and virtual photon are
parameterized as

$$p=(p_z,\vec{0},p_z)\equiv p_+\overline{n},$$
and
$$q\equiv q_+\overline{n}+q_-n \equiv -x_Bp_+\overline{n}+\frac{Q^2}{2x_Bp_+}n. \eqno(2.8)$$

    The contributions of the $q\rightarrow qg$ process in Fig. 3
to the coefficient function are

$$C_{CVPT}^{twist-2}$$
$$=\sum_i e^2_i\int \frac{d^4k}{(2\pi)^3}\frac{\delta((k+q)^2)\delta((p-k)^2)}{k^4}
\overline{M}(\gamma^*q\rightarrow \gamma^*q),\eqno(2.9)$$
where

$$ \overline{M}(\gamma^*q\rightarrow \gamma^*q)$$
$$=g^2<\frac{4}{3}>_{colour}\frac{1}{4}Tr[\gamma_{\mu}\gamma\cdot{k}_1
\gamma_{\nu}\gamma\cdot{k}\gamma_{\beta}\gamma\cdot{p}\gamma_{\alpha}\gamma\cdot{k}]
\Gamma_{\alpha\beta}(k_2)d_{\perp}^{\mu\nu}. \eqno(2.10)$$

$\Gamma_{\alpha\beta}$ is an axial gauge gluon polarization

$$\Gamma_{\mu\nu}(k)=g_{\mu\nu}-\frac{k_\mu{n}_\nu+k_\nu{n}_\mu}{k\cdot{n}}.  \eqno(2.11)$$

 In the calculations we use the Sudakov variables

$$k_{\mu}=bp_{\mu}+cq_{\mu}^\prime+k_{\perp\mu}, \eqno(2.12)$$
where two parameters are determined by the on-shell conditions

$$b=x_B-(1-x_B)\frac{Q^2}{2p\cdot q},\eqno(2.13)$$
and

$$c=\frac{Q^2}{2p\cdot q}. \eqno(2.14)$$

The result in the leading logarithmic region, where $k^2\ll Q^2$
are

$$C_{CVPT}^{twist-2}=\sum_i e^2_i \frac{\alpha_s}{2\pi}\int
\frac{dk^2_{\perp}}{k^2_{\perp}}\frac{4}{3}\frac{1+x_B^2}{1-x_B}.
\eqno(2.15)$$

The splitting function Eq. (2.7) can be directly obtained through
dividing the coefficient function $C_{CVPT}^{twist-2}$ by the
contributions of a bare photon-quark vertex (see Fig. 3b), which
is

$$C_{bare}^q=\sum
e^2_i\delta((q+p)^2)\overline{M}(\gamma^*q\rightarrow \gamma^*q)$$
$$=\sum e^2_i. \eqno(2.16)$$
That is

$$P_{CVPT}^{q\rightarrow q}\equiv \frac{C_{CVPT}^{twist-2}}{C_{bare}^q}=
\frac{4}{3}\frac{1+z^2}{1-z}, \eqno(2.17)$$ in which we have taken
the assumption that

$$p_+\gg q_-, \eqno(2.18)$$
thus, one can define

$$k_+=xP_+. \eqno(2.19)$$.

   It is not surprising of the above equivalence of two approaches.
In fact, the contributions of the backward components in two
off-shell Feynman propagators in Fig. 3 are suppressed under the
conditions Eq. (2.18) and $LLA(Q^2)$, although straightforward
computations in covariant theory do not display this fact. For
confirming this conclusion, we recalculate Eq.(2.15) but removing
the backward components in two Feynman propagators with the
momentum $k$ in Fig. 3.  We get the same result as Eq. (2.17).

\newpage
\begin{center}
\section{Twist-4 recombination functions}
\end{center}

    The recombination functions in the twist-4 evolution equations at
$LL(Q^2)A$ are depicted in Fig. 1b. A simplest way of the
derivation of the recombination functions is TOPT, which was
developed in [4]. In this method the recombination function is
isolated from the twist-4 coefficient function using the
equivalent approximation as shown in Fig. 4 and it is generally
written as

$$\alpha_s^2R_{TOPT}^{(p_1p_2\rightarrow \hat{k})}dx_4\frac{dk_\perp^2}{k_\perp^4}
=\frac{1}{16\pi^2}\frac{x_3x_4}{(x_1+x_2)^3}|M_{p_1p_2\rightarrow
\hat{k}k_2}|^2dx_4\frac{dk_\perp^2}{k_\perp^4}, \eqno(3.1)$$ where
$k_\perp^2=k_x^2+k_y^2$ and we use a current (the dashed lines)
$-\frac{1}{4}F^{i\mu\nu}F^i_{\mu\nu}$ to probe the gluonic current
[8].

    The momenta of the initial and final partons are
parameterized as

$$p_1=[x_1p,0,0,x_1p],~  p_2=[x_2p,0,0,x_2p],$$
$$\hat{k}=[x_3p+\frac{k_{\perp}^2}{2x_3p},k_{\perp},x_3p],~
k_2=[x_4p+\frac{k_{\perp}^2}{2x_4p},-k_{\perp},x_4p],$$
$$k_3=[(x_4-x_2)p+\frac{k_{\perp}^2}{2x_4p},-k_{\perp},(x_4-x_2)p],~$$
$$k_4=[(x_4-x_2)p+\frac{k_{\perp}^2}{2x_4p},-k_{\perp},(x_4-x_2)p]. \eqno(3.2)$$

    For example, the recombination function for $GG\rightarrow G$
at $t$-channel is

$$R^{GG\rightarrow G}_{TOPT}=\frac{g^4}{4}
{\langle\frac{9}{8}\rangle}_{color}\frac{x_3x_4}{(x_1+x_2)^3}
C_{\alpha\lambda\rho}C_{\beta\eta\sigma}
C_{\alpha^{\prime}\lambda^{\prime}\rho^{\prime}}
C_{\beta^{\prime}\eta^{\prime}\sigma^{\prime}}
d_{\perp}^{\rho\rho^{\prime}} d_{\perp}^{\sigma\sigma^{\prime}}$$
$$\times \frac{\Gamma_{\lambda\beta}(k_3)
\Gamma_{\lambda^{\prime}\beta^{\prime}}(k_4)}{k_3^2k_4^2}
\left(\delta^{lm}-\frac{k_3^lk_3^m}{|\overrightarrow{k_3}|^2}\right)
\left(\delta^{rs}-\frac{k_4^rk_4^s}{|\overrightarrow{k_4}|^2}\right),\eqno(3.3)$$
where $l,m$, $r, s$ are the space indices of $\alpha, \eta$,
$\alpha^{\prime}, \eta^{\prime}$ of $k_3$ and $k_4$, respectively;
$C_{\alpha\lambda\rho}$ and $C_{\beta\eta\sigma}$ are triple gluon
vertex. A set of recombination functions are listed in Table I.

    Now we return to discuss the recombination
functions in CVPT.  The gluon recombination functions in the
GLR-MQ evolution equation are derived using CVPT at the double
leading logarithmic approximation ($DLLA$), i.e., at the small $x$
limit [2]. We use the same way to recompute the recombination
functions but in whole $x$ region.

     The coefficient function containing the gluon recombination function
can be written as
$$C^{GG\rightarrow G}_{CVPT}=\int\frac{d^3k}{(2\pi)^3}{\delta(k_1^2)
\delta(k_2^2)}|M_{GG\rightarrow G}|^2. \eqno(3.4)$$

    In the $t$-channel,

$$|M_{GG\rightarrow G}|^2=\frac{g^4}{4}{\langle\frac{9}{8}\rangle}_{color}
C^{\alpha\rho\lambda}C^{\beta\eta\sigma}
C^{\alpha^{\prime}\rho^{\prime}\lambda^{\prime}}
C^{\beta^{\prime}\eta^{\prime}\sigma^{\prime}}
v^{\mu\mu^{\prime}}v^{\nu\nu^{\prime}}d^{\rho\rho^{\prime}}_{\perp}
d^{\sigma\sigma^{\prime}}_{\perp}$$
$$\frac{\Gamma_{\lambda\beta}(k_3)\Gamma_{\lambda^{\prime}\beta^{\prime}}(k_4)
\Gamma_{\mu^{\prime}\nu^{\prime}}(k_1)\Gamma_{\mu\alpha}(k)
\Gamma_{\nu\alpha^{\prime}}(k)\Gamma_{\eta\eta^{\prime}}(k_2)}
{k^4k_3^2k_4^2}$$ where the current-gluon vertex $v$ is
$$v^{\mu\nu}=k_\nu k_{1\mu}-g_{\mu\nu}k\cdot k_1,\eqno(3.5)$$
and

$$\Gamma_{\lambda\beta}(k_3)=g_{\lambda\beta}-
\frac{k_{3\lambda}n_{\beta}+k_{3\beta}n_{\lambda}}{k_3\cdot n}
=d^{\lambda\beta}_{\perp}-2\frac{k_3^-}{k_3^+}n^{\lambda}n^{\beta}.\eqno(3.6)$$

    The coefficient function of the contributions of a
bare gluonic vertex is

$$C_{bare}^G=\frac{x_B}{2Q^2}. \eqno(3.7)$$
Thus, the recombination function

$$R_{CVPT}^{GG\rightarrow G}\equiv \frac{C^{GG\rightarrow
G}_{CVPT}}{C_{bare}^G}. \eqno(3.8)$$

    The set of recombination functions using the above mentioned
CVPT method are given in Table II. A surprising result is that the
recombination functions with the IR singularities in the CVPT
method are inconsistent with that in the TOPT method. We shall
detail it in the following section.

\newpage
\begin{center}
\section{IR safety in the recombination functions}
\end{center}

    At first step, we point out that the IR singularities in
$R_{CVPT}$ originate from the gauge term in the gluon propagator
$\Gamma^{\lambda\beta}=
g^{\lambda\beta}-(k_3^{\lambda}n^{\beta}+k_3^{\beta}
n^{\lambda})/k_3\cdot n$, where $k_3\cdot n\sim y-x$ arris
divergence if $y\rightarrow x$.

    We decompose the propagator with the momentum $k$ on the two
gluon legs according to TOPT in Fig. 5. The on-shell propagating
momenta are

$$k\rightarrow \hat {k}_F(zp_+\overline{n}, \frac{k_{\perp}^2}{2zp_+}n, k_{\perp}), \eqno(4.1)$$
and

$$ \hat {k}_B(\frac{k_{\perp}^2}{2zp_+}\overline {n},zp_+n,  k_{\perp}). \eqno(4.2) $$
Where the sub-indices $F$ and $B$ refer to the forward and
backward propagations, respectively. The dominant contributions in
the $LLA(Q^2)$ are from those terms with the highest power of
$k_{\perp}^2$ in the numerator of the propagator. Obviously, terms
$n^{\lambda}$ or $d^{\lambda\beta}_{\perp}$ in Eq. (3.6) can
combine with $\hat {k}_B$ or $\hat{k}_F$ in Eqs. (4.1) and (4.2)
through the quark-gluon or triple-gluon vertex, respectively. In
consequence, the contributions from the forward and backward
components are coexist in the two propagators with the momentum
$k$ in these diagrams.

    It is different from the recombination functions $R_{CVPT}^{GG\rightarrow G}$, the
dominant numerator factor  $\gamma\cdot \overline{n}$ in
$R_{CVPT}^{GG\rightarrow q}$ only chooses $\hat{k}_F$ in Eq.
(4.2). Therefore, the contributions from the backward components
are suppressed at the $LLA(Q^2)$ and the equivalent particle
approximation is available as in $R_{TOPT}^{GG\rightarrow q}$.

    Such gauge singularities also exist in the
derivation of the GLR-MQ equation [2], where an unusual
$i\epsilon$ prescription is used at limit $x\ll 1$ and the result
is finite after taking the principal value. Obviously, the
contributions from the backward components in the gluon
propagators can not been entirely excluded in this way. Thus, we
can conclude that the recombination functions in the GLR-MQ
equation are really a part of the twist-4 coefficient functions
but not the evolution kernels for parton distribution
functions(PDFs) since two legs in Fig. 5 are off-shell.

    Of cause, any physical results are IR safe. The IR divergences
should be cancelled by summing up all related diagrams, including
interference and virtual diagrams (see Fig. 6). After that one can
isolate the recombination functions from the coefficient functions
in the covariant framework. However, at moment there is no
available covariant way to calculate the virtual and interference
diagrams like Fig. 6. In the GLR-MQ equation such summations are
evaluated using the Abramowsky-Gribove-Kancheli (AGK) cutting rule
[9]. However, this application of the AGK cutting rule has been
argued in Ref. [4]. Besides, the AGK cutting rules only change a
relative weight in the contributions of the real diagrams and they
can not cancel any IR divergences.

    Fortunately, the TOPT approach in the Bjorken frame provides
an available method to derive the recombination functions. In
fact, the recombination functions can be safely separated from the
twist-4 coefficient functions in this way, since the backward
components in two parton legs with the momentum $k$ are
suppressed. Thus, the equivalent parton approximation can be used.
Furthermore, the contributions of the gauge singular terms
disappear due to the absence of these backward components. For
justifying this conclusion, we use $d_{\perp}^{\mu\nu}$ to replace
$\Gamma^{\mu\nu}$ in Eq. (3.3) and get the same results as Table
I.

    In the case that all propagators only keep their forward
component in Fig. 6, the calculations of the relating interference
and virtual diagrams become possible. In fact, a simple relation
among the real-, virtual- and interference diagrams is derived in
Ref. [4]. Thus, we can pick up the contributions of leading
recombination from all necessary cut diagrams. The resulting
evolution kernel for the gluon recombination function has
following form [4]:

$$\int _{2(x_1+x_2)>x_B}dx_1dx_2f(x_1,x_2,Q^2)R_{TOPT}^{GG\rightarrow G}(x_1,x_2, x_B)$$
$$-2\int_{(x_1+x_2)>x_B}dx_1dx_2f(x_1,x_2,Q^2)R_{TOPT}^{GG\rightarrow
G} (x_1,x_2,x_B), \eqno(4.3)$$ where $f(x_1,x_2,Q^2)$ is the two
gluon correlation function. The first and second terms in Eq.
(4.3) give the antishadowing and shadowing effects in the
evolutions of the parton densities.

    Now let us look back to the CVPT method for the derivation of
the recombination functions. According to the above mentioned
discussions, we use $d_{\perp}^{\lambda\beta}$ to instead
$\Gamma_{\lambda\beta}$ in Eq. (3.6). As expected, one can get the
same results as Table I. From the above mentioned discussions we
can conclude that the forward propagators dominate the the
recombination functions.

\newpage
\begin{center}
\section{Discussions and summary}
\end{center}

    An interesting question is why the IR singularities of
the gluon propagator in the twist-2 coefficient function (Fig. 3a)
does not break the equivalent particle approximation.  We note
that this propagator through the dashed line is on-shell. Thus,
Eq. (2.11) can be rewritten as

$$\hat{\Gamma}^{\alpha\beta}=d_{\perp}^{\alpha\beta}-
\frac{k_{\perp}^2}{k_{2+}^2}n^\alpha n^\beta\eqno(5.1),$$ Now the
dominant contributions to the $LL(Q^2)A$ are only from the forward
components and the equivalent particle approximation is
applicable.

    Our discussions for the recombination functions can also be
used in the investigation of the twist-4 coefficient functions
[11]. One can get those functions up to $dk^2_{\perp}/k^4_{\perp}$
accuracy by the product of the recombination functions in Table I
and the bare vertex (2.16) or (3.7).

    In summary, we show that the QCD evolution kernels are
frame- and gauge-independent, however, the separation of these
kernels from the coefficient functions depend on the frame and
gauge. The IR singularities in the recombination functions inhibit
us to safely isolate the recombination functions. This result
leads to the fact that the gluon recombination functions in the
GLR-MQ evolution equation are unavailable. The methods avoiding
the IR divergences are discussed by using the TOPT.

\noindent {\bf Acknowledgments}:
 This work was supported by the National Natural Science Foundations of China 10075020,
 90103013 and 10135060.

\newpage
\begin{center}Table.1

\begin{tabular}
{|l|l|}\hline$R^{GG\rightarrow{q}}_{TOPT}
=R^{GG\rightarrow{\bar{q}}}_{TOPT}
$&${\frac
{1}{96}}\,{\frac { \left( 2\,y-x \right) ^{2}
 \left(18\,{y}^{2}-21\,yx+ 14\,{x}^{2}
 \right)}{{y}^{5}}}$\\\hline
 $R^{qq\rightarrow{q}}_{TOPT}$&$\frac{2}{9}{\frac
{ \left( 2\,y-x \right) ^{2}}{{y}^{3}}}$\\\hline
$R^{q\bar{q}\rightarrow{G}}_{TOPT}$&${\frac{4}{27}}\,{\frac {
\left( 2\,y-x \right) \left(18\,{y}^{2}-9\,yx + 4\,{x}^{2}\right)
}{{y}^{3}x}} $\\\hline $R^{GG\rightarrow{G}}_{TOPT}$&${\frac
{9}{64}}\,{\frac { \left( 2\,y-x \right)  \left(
72\,{y}^{4}-48\,{y}^{3}x+140\,{y}^{2}{x}
^{2}-116\,y{x}^{3}+29\,{x}^{4} \right) }{{y}^{5}x}}$\\\hline
\end{tabular}
\end{center}

\begin{center}Table.2

\begin{tabular}
{|l|l|}\hline $R^{GG\rightarrow{q}}_{CVPT}
=R^{GG\rightarrow{\bar{q}}}_{CVPT}
$&${\frac
{1}{96}}\,{\frac { \left( 2\,y-x \right) ^{2} \left(
18\,{y}^{2}-21\,yx+14\,{x}^{2} \right)}{{y}^{5}}}$ \\\hline
$R^{qq\rightarrow{q}}_{CVPT}$&${\frac
{2}{9}}\,{\frac { \left( 2\,y-x \right) ^{2} \left( 13\,{y}^{2
}+6\,yx+3\,{x}^{2} \right) }{{y}^{3} \left( y-x \right)
^{2}}}$\\\hline
$R^{q\bar{q}\rightarrow{G}}_{CVPT}$&${\frac {4}{27}}\,{\frac {
\left( 2\,y-x \right) \left(18\,{y}^{2} -9\,yx+4\,{x}^{2} \right)
}{{y}^{3}x}} $\\\hline $R^{GG\rightarrow{G}}_{CVPT}$&${\frac
{9}{64}}\,{\frac { \left( 2\,y-x \right)
\left(144\,{y}^{6}-360\,{y}^{5}x+541\,{y}^{4}{x}^{
2}-440\,{y}^{3}{x}^{3}+221\,{y}^{2}{x}^{4} -
70\,y{x}^{5}+12\,{x}^{6} \right) }{{y}^{5} \left( y-x \right) ^{
2}x}}$\\\hline

\end{tabular}
\end{center}

\newpage

\noindent {\bf Figure Captions}

\noindent Fig. 1  The schematic descriptions of (a) splitting and
(b) recombination functions.  The solid lines are gluon or quarks
and the grey ovals imply all possible QCD channels.\\

\noindent Fig. 2  Graphical illustration of the isolation of a
splitting function in the TOPT cut diagram. $\hat {k}$ is on-shell
momentum and dashed lines refer the time order.\\

\noindent Fig. 3  (a) A dominate covariant Feynman diagram for the
twist-2 coefficient function at the $LLA$; (b) the bare virtual photon-quark vertex.\\

\noindent Fig. 4  Graphical illustration of the isolation of a
recombination function in the TOPT cut diagram.The dashed line is
an effective "current" probing the gluonic matrix. \\

\noindent Fig. 5 A covariant Feynman diagram containing the gluon
recombination function.

\noindent Fig. 6  The virtual and interferant diagrams relating to
the gluon recombination function.


\newpage
\begin{thebibliography}{99}


\bibitem {1} L.V. Gribov, E.M. Levin and M.G. Ryskin, Phys. Rep. 100 (1983)
1.

\bibitem {2} A.H. Mueller and J. Qiu, Nucl. Phys. B268, 427 (1986).

\bibitem {3} A. H. Mueller, Nucl. Phys. B 335, 115 (1990); A. H. Mueller, Nucl. Phys. B 415, 373 (1994).

\bibitem {4} W. Zhu, Nucl. Phys. B 551, 245 (1999) [hep-ph/9809391].

\bibitem {5} G. Sterman, G., 1993 An introduction to Quantum Field Theory, (Cambridge
University Pres), 266 (1993); M.D. Scadron, Advanced Quantum
theory and Its Applicatiopns Through Feynman Diagrams, eds W.
Beigb$\ddot{o}$ck, R.P. Geroch, E.H. Lieb, T. Regge and W.
Thirring (Springer-Verlag, Berlin Heidelberg), 159 (1979).

\bibitem {6} G. Altarelli and G. Parisi, Nucl. Phys. B 126, 298 (1977).

\bibitem {7} P. Kessler, Nuovo Comento 16, 809 (1966); V.N Baier, V.S. Fadin and V.A. Khoze,  Nucl. Phys. 65, 381 (1973);

M.S. Chen and P. Zerwas, Phys. Rev. D12, 187 (1975).

\bibitem {8} A. Mueller, Phys. Rep. 73, 237 (1981).

\bibitem{10} V.A. Abramovsky, J.N. Gribov and O.V. Kancheli, Sov.
J. Nucl. Phys. $\bf 18$, 593 (1973).

\bibitem{11} J. Bl\"{u}mlein, V. Ravindran, J.H. Ruan and W. Zhu,
Phys. Lett. B 504, 235 (2001).


\end{thebibliography}
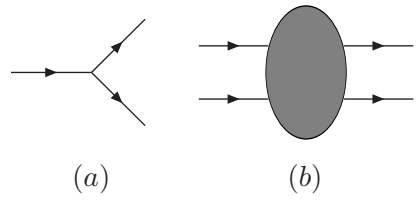
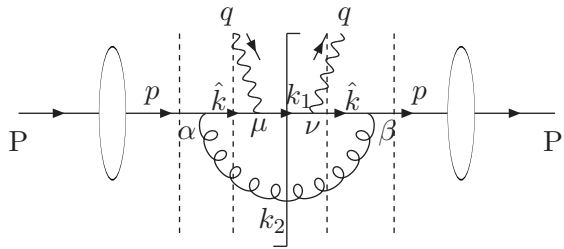
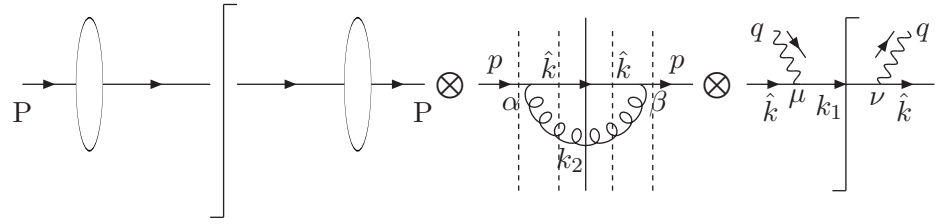
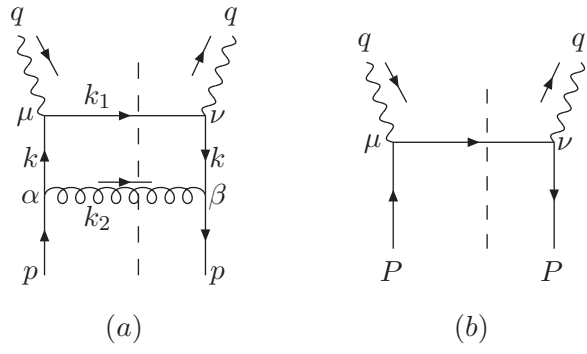
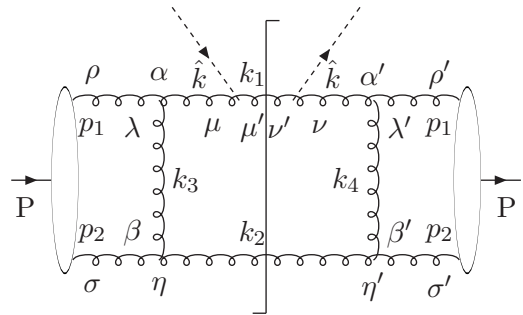
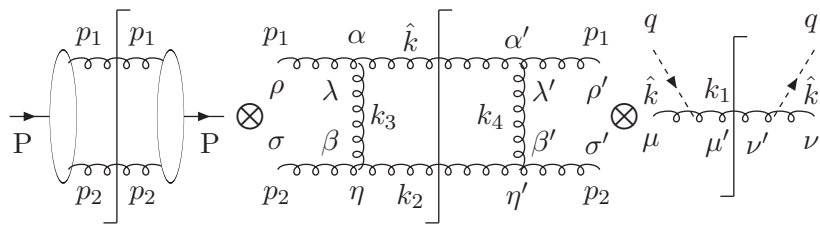
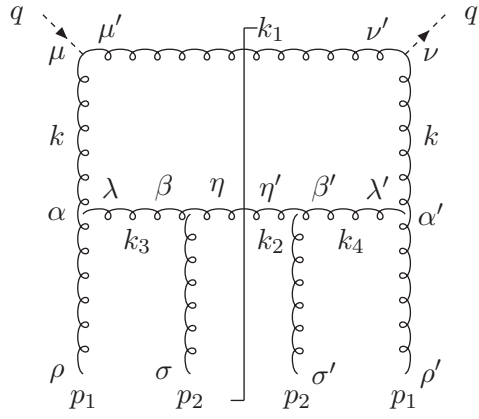
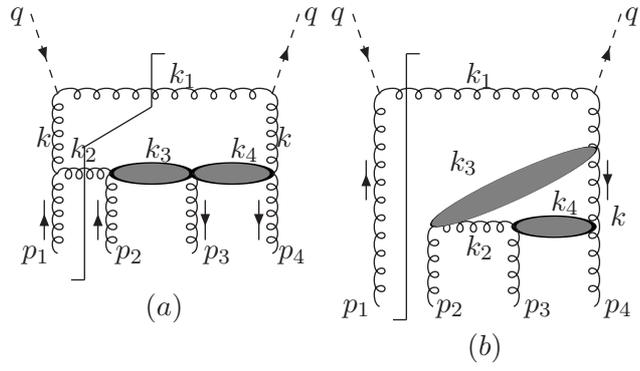
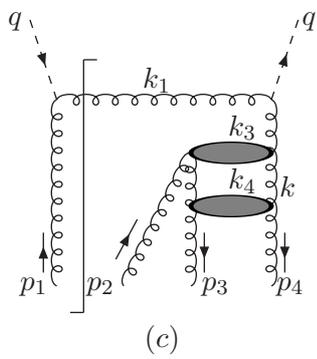
\end{document}